\documentclass[12pt]{article}
\usepackage{amsfonts,pifont}
\usepackage{amsmath,chemarrow}
\usepackage{amssymb}
\usepackage{amstext}
\usepackage{amsfonts}
\usepackage{graphicx}
\usepackage{indentfirst}
\usepackage{hyperref,comment}

\makeatletter
\def\ExtendSymbol#1#2#3#4#5{\ext@arrow 0099{\arrowfill@#1#2#3}{#4}{#5}}
\def\RightExtendSymbol#1#2#3#4#5{\ext@arrow 0359{\arrowfill@#1#2#3}{#4}{#5}}
\def\LeftExtendSymbol#1#2#3#4#5{\ext@arrow 6095{\arrowfill@#1#2#3}{#4}{#5}}
\makeatother

\begin{document}
\baselineskip 20pt

\title{Generation of Einstein-Podolsky-Rosen State via Earth's Gravitational Field}
\author{Jun-Li Li$^1$ and
Cong-Feng Qiao$^{1,2}$\footnote{Corresponding author, qiaocf@gucas.ac.cn.}\\[0.5cm]
$^{1}$ Department of Physics, University of Chinese Academy of Sciences \\
YuQuan Road 19A, Beijing 100049, China\\[0.2cm]
$^{2}$Theoretical Physics Center for Science Facilities (TPCSF),
CAS\\ YuQuan Road 19B, Beijing 100049, China}

\date{}
\maketitle

\begin{abstract}
Although various physical systems have been explored to produce
entangled states involving electromagnetic, strong, and weak
interactions, the gravity has not yet been touched in practical
entanglement generation. Here, we propose an experimentally feasible
scheme for generating spin entangled neutron pairs via the Earth's
gravitational field, whose productivity can be one pair in every few
seconds with the current technology. The scheme is realized by
passing two neutrons through a specific rectangular cavity, where the
gravity adjusts the neutrons into entangled state. This provides a
simple and practical way for the implementation of the test of
quantum nonlocality and statistics in gravitational field.

\end{abstract}

Ever since the discovery of Bell inequalities \cite{JSBell}, the
generation of entanglement with various physical systems has been the
intensively studied subject. Now the entangled states can be
generated not only in optical \cite{photon-Aspect,photon-Weihs},
atomic \cite{ions-Rowe}, solid state \cite{Josephson-Ansmann} systems
where only the electromagnetic interactions is involved, but also
with baryons \cite{Tornqvist-Baryon}, leptons \cite{ee-tau}, and
mesons \cite{bertlmann-rev, three-intactions, tau-charm-factory,
new-possibility} where the strong or weak interaction emerges as the
dominant force. Many of these entanglement generation schemes have
been experimentally realized in testing the violation of Bell's
inequalities which reveal the nonlocality of quantum theories, e.g.
\cite{photon-Aspect,photon-Weihs, Tornqvist-Baryon} etc. Some of the
systems have further found their roles in quantum computations and
quantum information processing, see \cite{Solit-computation,
Ion-computation, Photon-computation} and references therein. Besides
these practical contributions as a crucial physical resource for
quantum information science, one considerable interest of exploring
these various entangled systems is to show that the nonlocal
correlation is not a peculiarity attributed to specific interactions
but a universal quantum phenomena.

It has been noticed that three of the four fundamental interactions
in nature, i.e., electromagnetic, strong, and weak, are capable of
generating entanglement leaving the gravity a sole exception
\cite{three-intactions}. Although some quantum effects of the
classical gravity have already been observed, i.e., quantum
interference \cite{N-Gravity-Interference,
proper-time-interferometric}, discrete energy levels of neutrons in
the gravitational potential \cite{N-Elevel-observe}, there is still
no report on quantum entanglement generation concerning the Earth's
gravitational filed. Despite being considered as the ideal tool to
test the Bell inequalities \cite{Neutron-P}, the extremely small
neutron-neutron scattering length ($\sim10^{-14}$ m \cite{N-Length})
makes the generation of entangled neutron pairs via their low energy
scattering a considerable technical challenge.  However, the
possibility of entangling two neutrons by successive scattering from
a macroscopic sample is still under studying \cite{two-N-scattering}.
Up to now, only the entanglement among different degree of freedoms
of a single neutron has been experimentally realized, see
\cite{Neutron-nat, neutron-3degree, Leggett-N, GHZ-N-polarimetry} and
reference therein.

In this paper, we propose a scheme to generate the
Einstein-Podolsky-Rosen state via Earth's gravitational field. The
scheme composed of three different functional components: an energy
filter that monochromatizes the neutrons, a rectangular cavity (RC)
which entangles two neutrons, and the neutron polarization analyzers
for revealing the spin correlations. The main idea is to guide a pair
of monochromatic neutrons into the RC where they are enforced into
the same energy state. The spin singlet state should then be obtained
due to the antisymmetrization requirement of the indistinguishable
fermions. We design a practical experimental setup for our scheme, by
which we show that, with the current technology, the gravity dominant
entanglement generation and the nonlocality test with entangled
neutron pairs are within the experimental reach.

As been observed in \cite{N-Elevel-observe}, neutrons falling towards
a horizontal reflecting mirror will distributed discontinuously in
the vertical direction. Such a system can be described by the quantum
theory of a particle bouncing in the gravitational field above a
perfect mirror \cite{N-qmotion-Voronin}. The Schr\"odinger equation
governing the motion of the neutrons in the vertical dimension reads
\begin{eqnarray}
-\frac{\hbar^2}{2m}\frac{\mathrm{d}}{\mathrm{d}\xi^2}\phi(\xi) +
V(\xi) \phi(\xi) =
E\phi(\xi)\; , \; V(\xi) =\left\{\begin{array}{ll} mg\xi & \xi \geq 0 \\
\\ +\infty & \xi<0 \end{array}\right.\; .
\label{1-dimentional-equation}
\end{eqnarray}
Here $\xi$ is the height of neutron from the horizontal mirror, $m$
is the mass of neutron, $g$ is the acceleration constant near the
Earth's surface. Eq.(\ref{1-dimentional-equation}) can be solved and
shows discrete energy eigenstates \cite{search-gravity-levels}
\begin{eqnarray}
\phi_{n}(\xi) = \mathcal{N}_n\mathrm{Ai}(\xi
\frac{(2m^2g)^{1/3}}{\hbar^{2/3}} - E_n
\frac{2^{1/3}}{(mg^2\hbar^2)^{1/3}}) \; , \label{1-D-wave-function}
\end{eqnarray}
where $\mathrm{Ai}$ are Airy functions and $\mathcal{N}_{n}$ is a
normalization constant, $ E_{n} = \alpha_n \displaystyle
\frac{(mg^2\hbar^2)^{1/3}}{2^{1/3}} $ corresponds to the energy
eigenvalue of state $\phi_n$ with $\alpha_n$ being the $n$th zero of
Airy function $\mathrm{Ai}(-\alpha_n)=0$.

\begin{figure}\centering
\scalebox{0.25}{\includegraphics{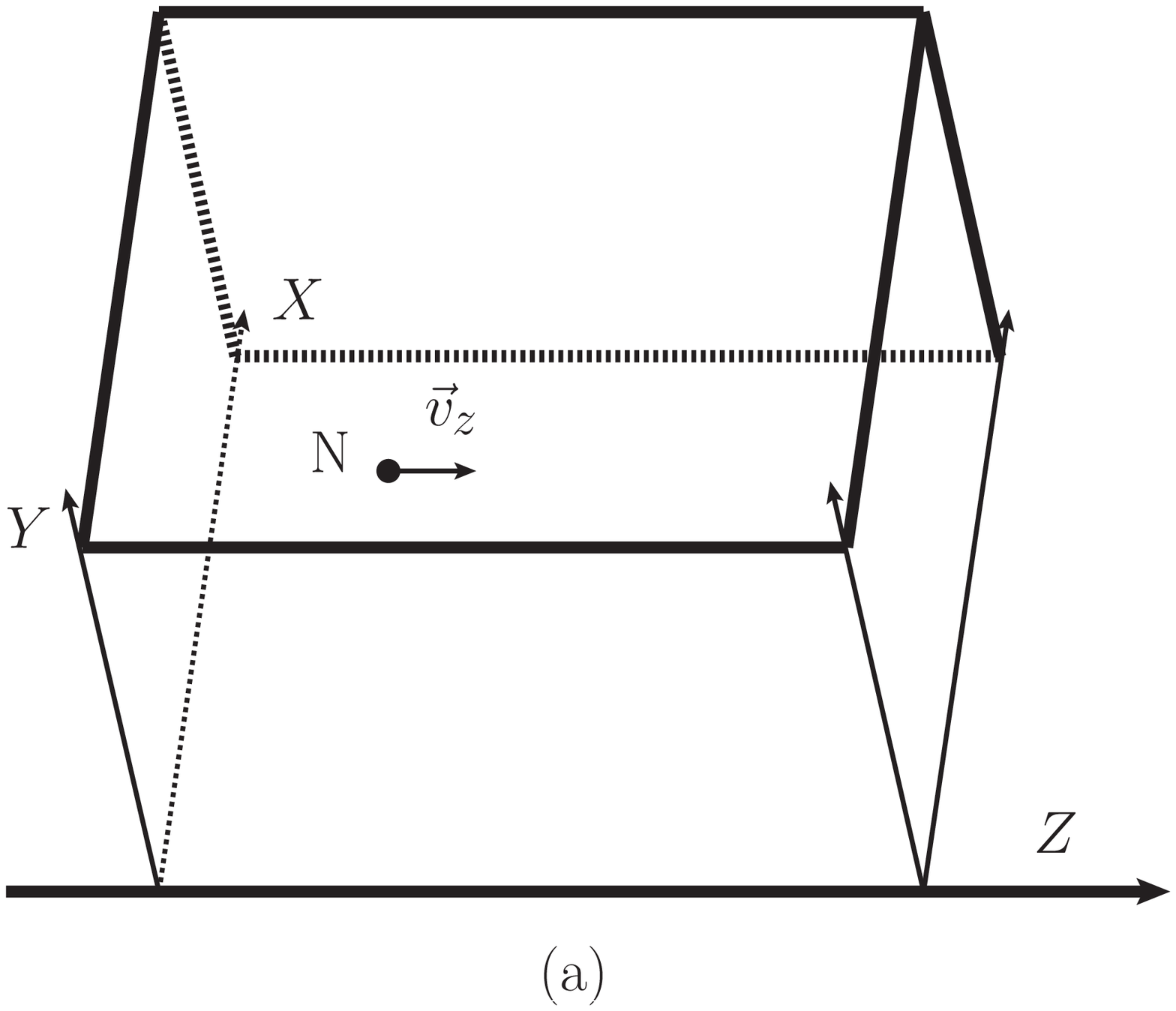}}\; \;
\scalebox{0.25}{\includegraphics{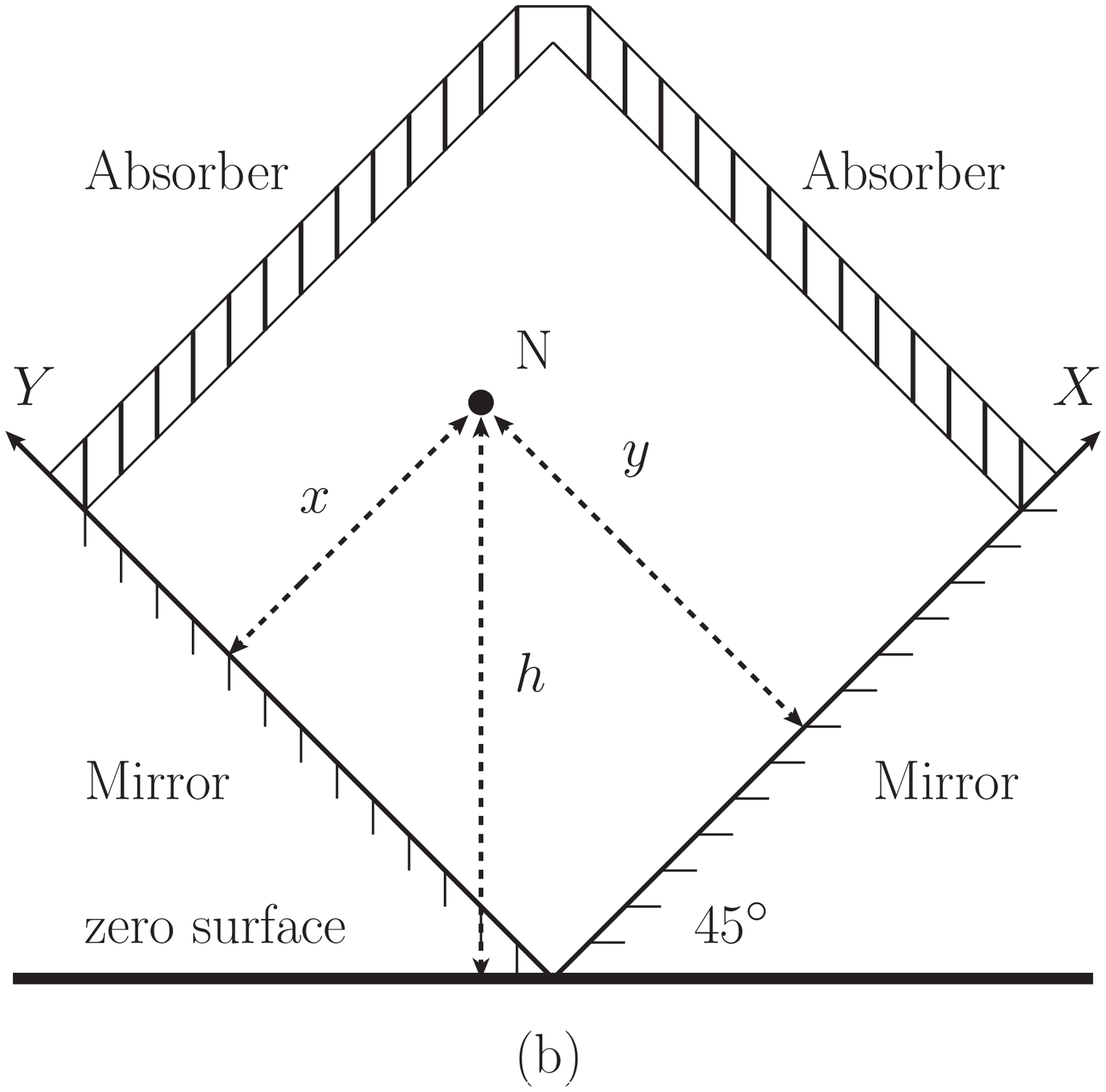}} \caption{
\small A rectangular cavity (RC) where the two lower surfaces are neutron
mirrors and the two upper surfaces are neutron absorbers. In (a) a
neutron N passes through the rectangular cavity with longitudinal
velocity $\vec{v}_z$. (b) represents the transection of the rectangular
cavity and $h$ is the height of the neutrons relative to the zero
gravitational potential surface.}\label{cavity+structure}
\end{figure}

\begin{figure}\centering
\scalebox{0.25}{\includegraphics{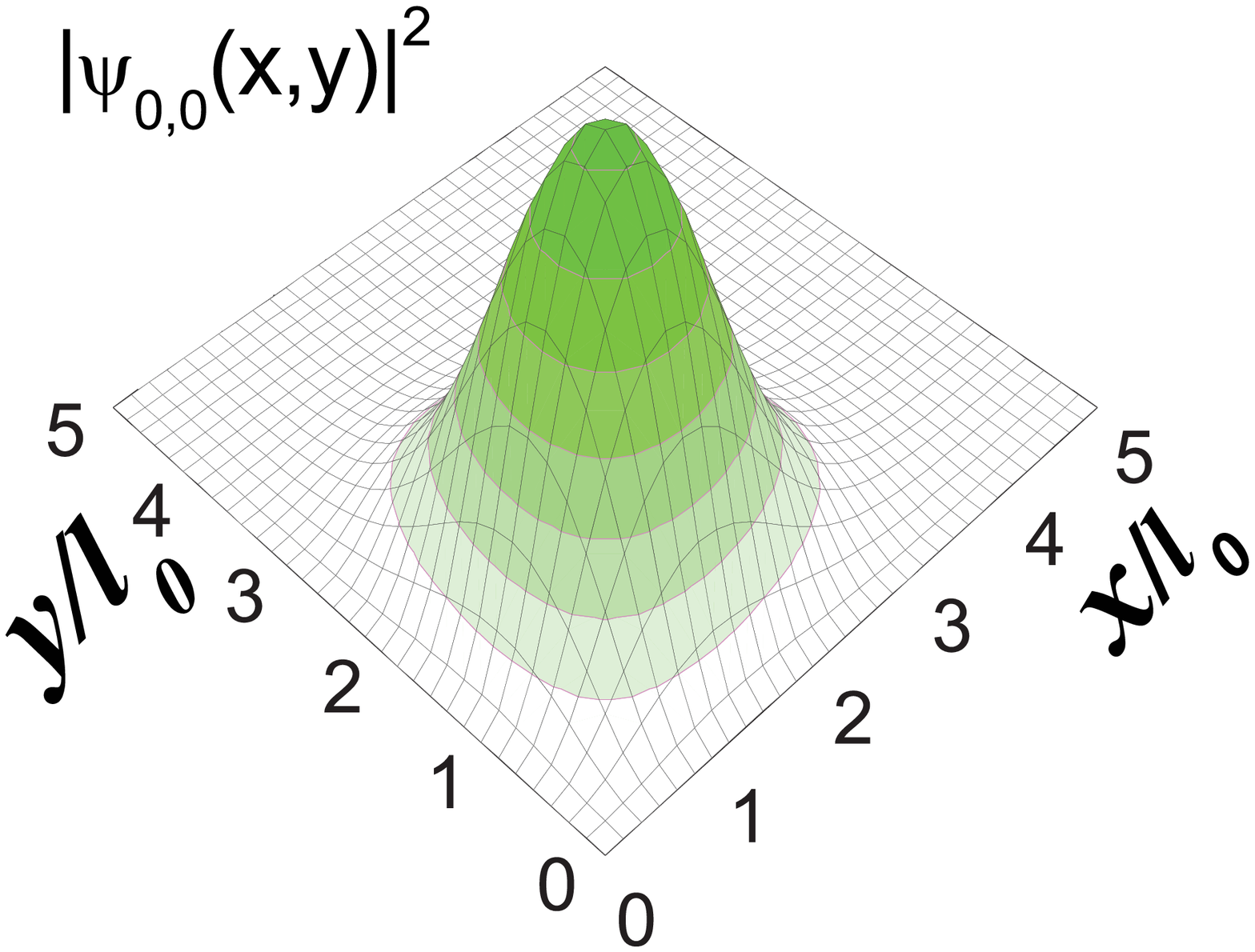}}\hspace{0.5cm}
\scalebox{0.25}{\includegraphics{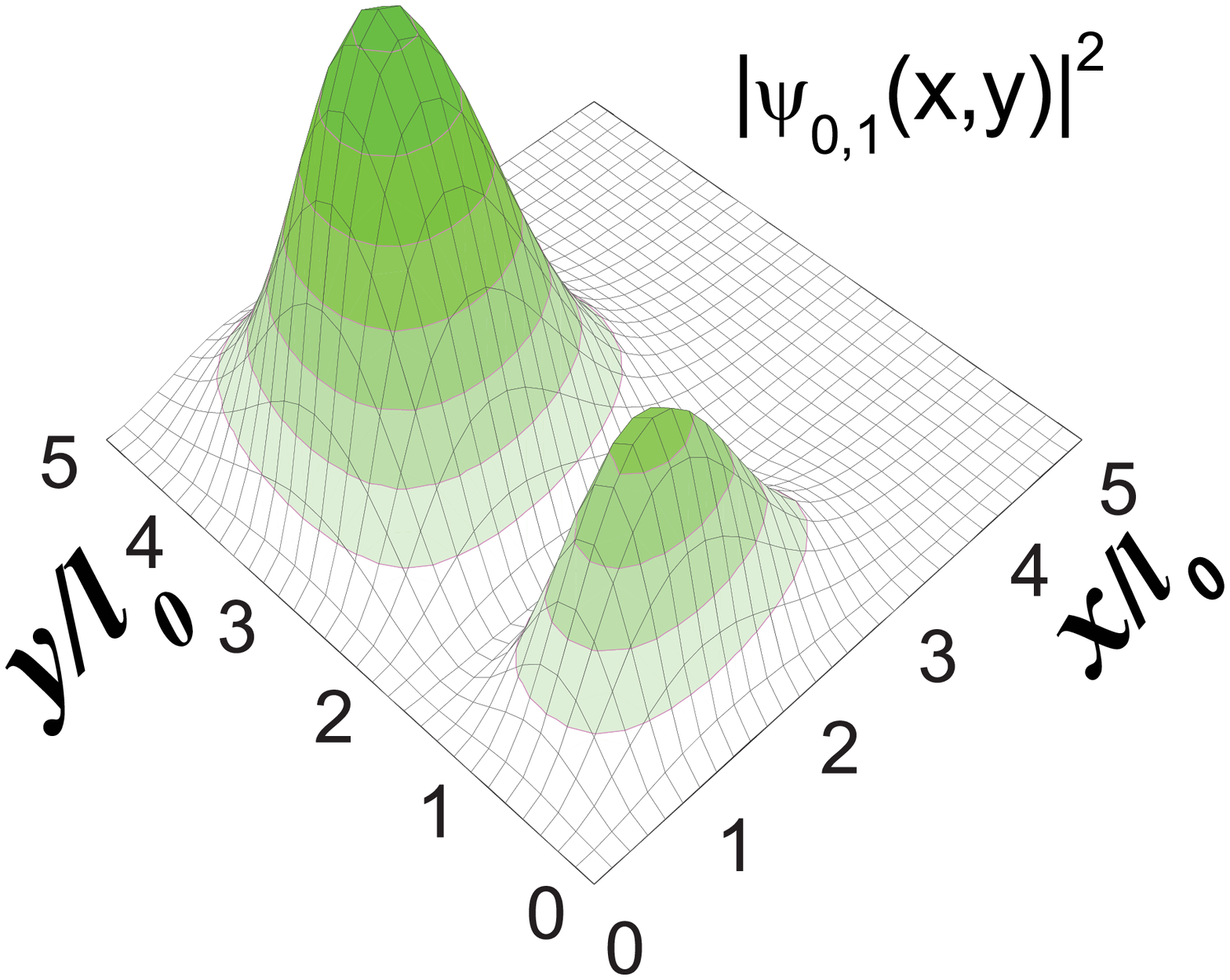}}\\ \vspace{0.5cm}
\scalebox{0.25}{\includegraphics{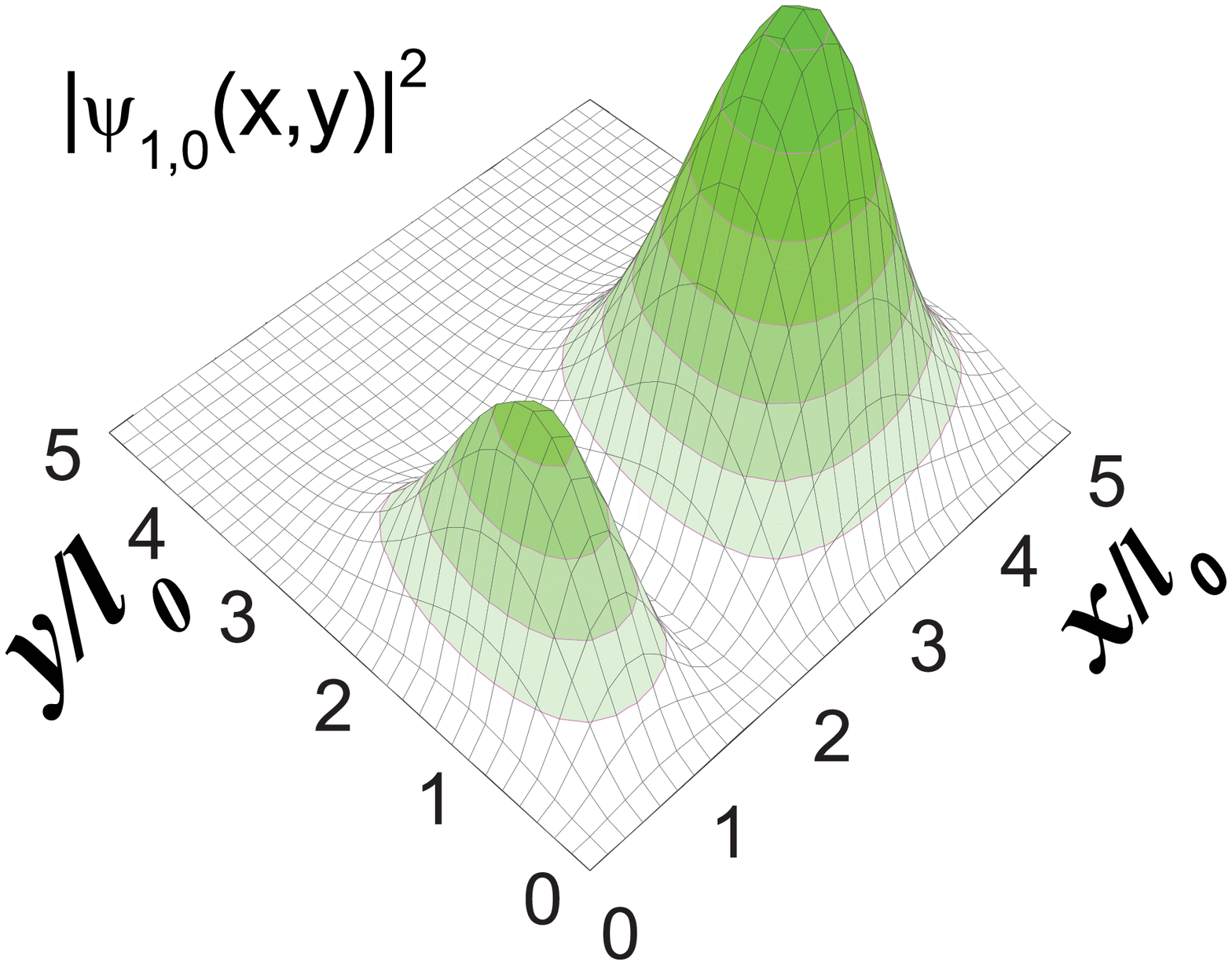}}\hspace{0.5cm}
\scalebox{0.25}{\includegraphics{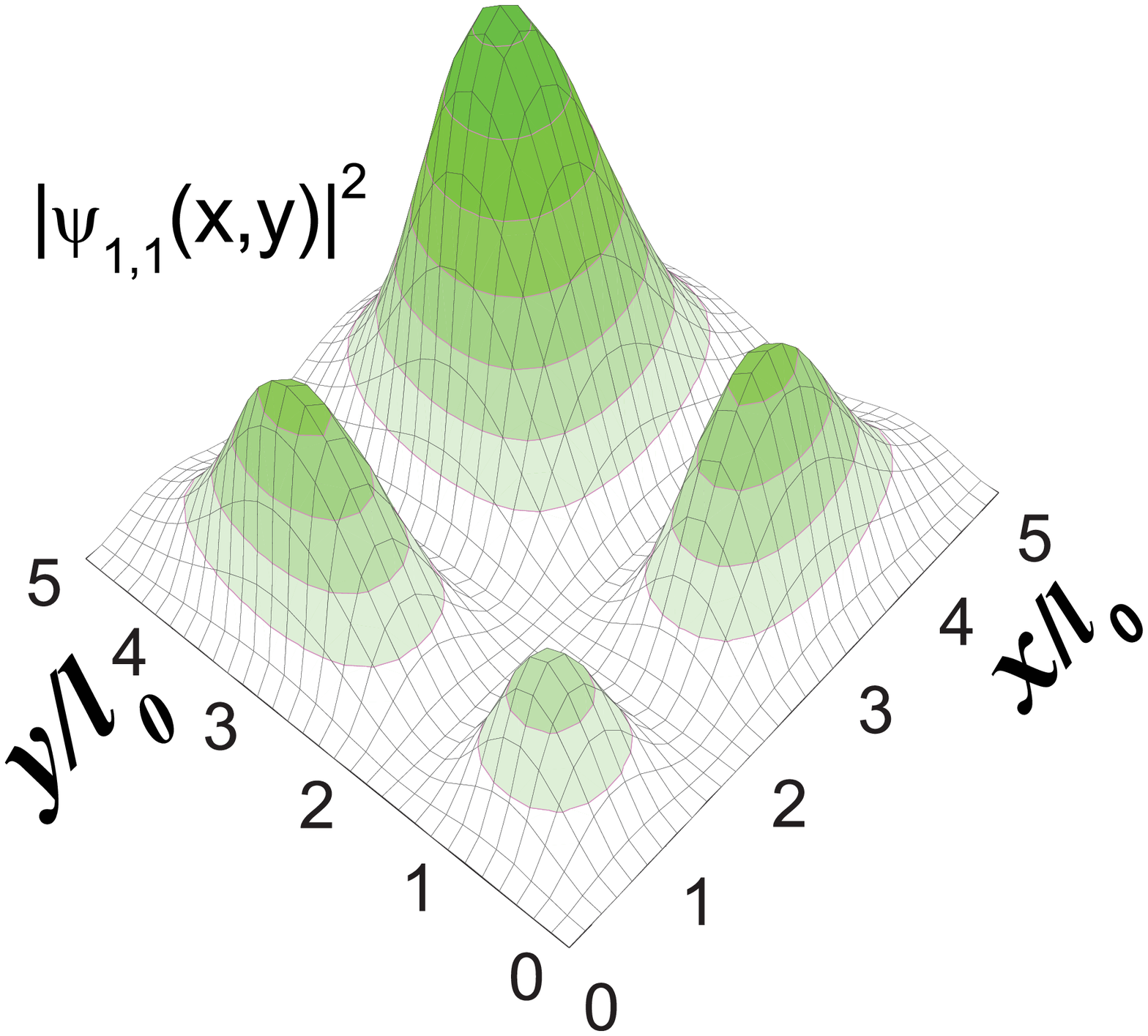}} \caption{ \small
The probability density distributions of the first four lowest eigenstates
in the transection of the rectangular cavity: $|\Psi_{0,0}|^2$,
$|\Psi_{0,1}|^2$, $|\Psi_{1,0}|^2$, $|\Psi_{1,1}|^2$. The $X$ and
$Y$ axes are in unit of $l_0$. } \label{N-wave-distribution}
\end{figure}

Now considering a specifically designed RC as shown in
Fig.\ref{cavity+structure}, where the two lower surfaces of the
cavity are neutron mirrors while the two upper surfaces are neutron
absorbers. For each neutron in this cavity, it is subjected to a
constant gravitational force $F = m g$. Together with the lower
mirrors, the Earth's gravitational field provides the confining
potential well for the neutrons, which is
\begin{eqnarray}
V(x,y) = \left\{\begin{array}{ll}mg(x+y)/\sqrt{2} & x\geq 0 \;
\text{and} \; y\geq 0
\\ \\ +\infty & x < 0 \; \mathrm{or}\;  y< 0\end{array}\right. \; ,
\label{V-Gravity-Potential}
\end{eqnarray}
where we have chosen the zero potential surface in Fig.
\ref{cavity+structure}. In the transection of the cavity, the
$X$-$Y$ plane of Fig. \ref{cavity+structure}(b), the Schr\"odinger
equation of motion for neutron takes the following form
\begin{eqnarray}
-\frac{\hbar^2}{2m} \left(\frac{\partial^2}{\partial
x^2}+\frac{\partial^2}{\partial y^2}\right)\Psi(x,y) + V(x,y)
\Psi(x,y) = E\, \Psi(x,y) \; . \label{N-wave-equation}
\end{eqnarray}
The energy eigenstates and eigenvalues of this equation can be
similarly obtained as that of Eq.(\ref{1-dimentional-equation}), and
can be simply formulated as
\begin{eqnarray}
\Psi_{n,m}(x,y) = \psi_n(x)\psi_m(y) \; , \; E_{n,m} = E_n+E_m \; .
\end{eqnarray}
Here, $\psi_{n}(x) = \mathcal{N}_n\,\mathrm{Ai}\left(x/l_0 -
E_{n}/\varepsilon_0 \right)$, $\mathcal{N}_n$ is the normalization
constant; $l_0$ and $\varepsilon_0$ are the characteristic length
and energy defined as
\begin{eqnarray}
l_0 & = & \hbar^{2/3}/(\sqrt{2}m^2g)^{1/3} \simeq 6.59 \cdot
10^{-6}\ \mathrm{m}\; ,  \\  \varepsilon_0 & = &
\sqrt[3]{mg^2\hbar^2/4} \simeq 4.78 \cdot 10^{-13} \ \mathrm{eV} \;
;
\end{eqnarray}
and $ E_{n} = \alpha_n \varepsilon_0$ is the eigen energy of $\psi_n$
with $\alpha_n$ being defined after equation
(\ref{1-D-wave-function}). The first four lowest eigenstates are
$\Psi_{0,0}$, $\Psi_{0,1}$, $\Psi_{1,0}$, and $\Psi_{1,1}$, with
$|\Psi_{n,m}(x,y)|^2$ being the value of the neutron probability
density distributions in the $X$-$Y$ plane, see Fig.
\ref{N-wave-distribution}. Their corresponding energies can be listed
as follows
\begin{eqnarray}
E_{0,0} & = &  E_0 + E_0 = \alpha_0 \varepsilon_0 + \alpha_0
\varepsilon_0\simeq 2.22 \times 10^{-12}\, \mathrm{eV}\; , \label{E00} \\
E_{0,1} & = & E_0 + E_1  = \alpha_0 \varepsilon_0 + \alpha_1
\varepsilon_0\simeq 3.06 \times 10^{-12} \, \mathrm{eV}\; , \label{E01}\\
E_{1,0} & = & E_1 + E_0  = \alpha_1 \varepsilon_0 + \alpha_0
\varepsilon_0\simeq 3.06 \times 10^{-12} \, \mathrm{eV}\; , \\
E_{1,1} & = & E_1 + E_1 =  \alpha_1 \varepsilon_0 =\alpha_1
\varepsilon_0 \simeq 3.90 \times 10^{-12} \, \mathrm{eV}\; ,
\end{eqnarray}
where $E_{0,0}$ is the ground state energy and $E_{0,1}$ and
$E_{1,0}$ are energies of the first degenerated excited states.

From the probability density distributions $|\Psi_{n,m}(x,y)|^2$
plotted in Fig. \ref{N-wave-distribution}, we see that the quantum
states of higher energies become densely distributed in the area with
larger values of $x$, $y$. If the two upper absorbers are set at a
position of $x = y \simeq 3l_0$, the transection of RC would have the
size holding only the ground state. In this configuration, while the
excited states $\Psi_{0,1}$, $\Psi_{1,0}$, $\Psi_{1,1}$ and other
even higher excited states are substantially absorbed when the
neutron transmitting in the cavity, the ground state survives. The
minimal time $\tau_g$ while the neutron has to stay in the cavity to
resolve different quantum states is \cite{search-gravity-levels}
\begin{eqnarray}
\tau_g = \hbar/(E_{0,1}-E_{0,0}) \sim 7.84\cdot 10^{-4}\mathrm{s}
\; .
\end{eqnarray}
After a significantly longer time duration $T \gg \tau_g$ in the
cavity, a substantial absorbtion of the excited state will be
expected in the present configuration of $x = y \simeq 3l_0$.

\begin{figure}\centering
\scalebox{0.5}{\includegraphics{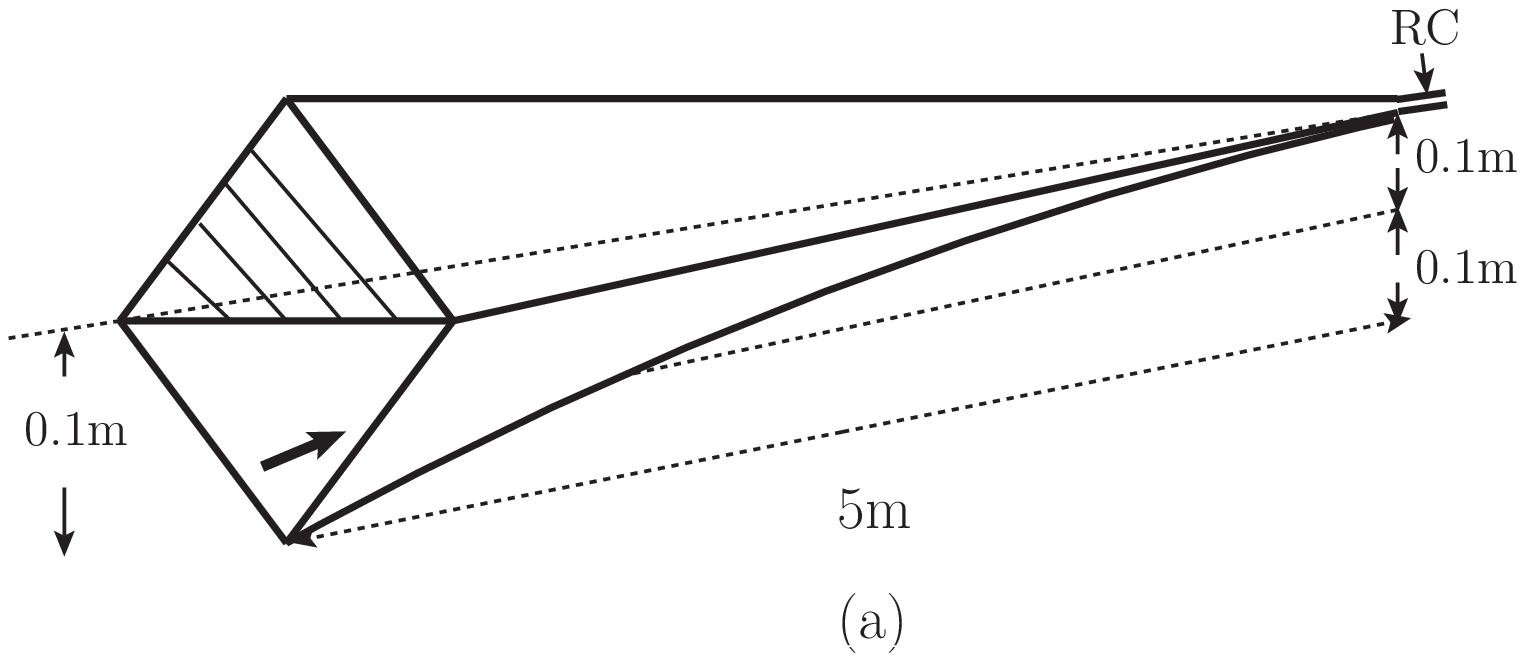}}
\scalebox{0.6}{\includegraphics{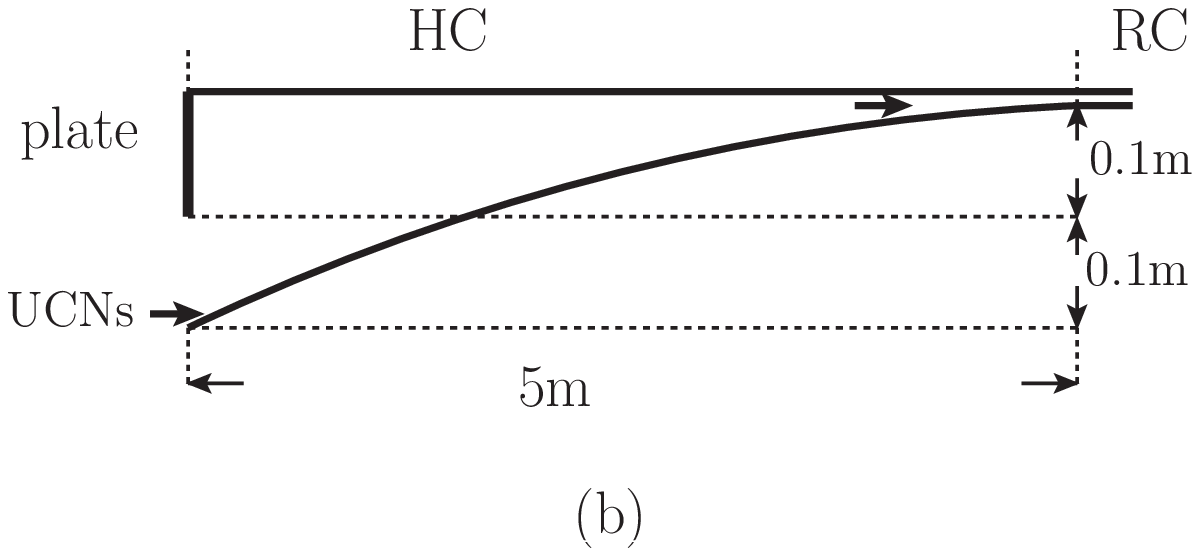}}
\caption{ An HC coupled to the RC (a) and
its longitudinal section (b). The whole HC
can be made up of neutron mirrors. The arc in (b) is part of a circle
with radius $62.6$m and is tangent
to the horizontal RC at the joint. The UCN beam ejected into the HC from
the lower half of the square (unshaded region in the (a)).}
\label{slope-cavity}
\end{figure}

Now suppose a beam of monochromatic UCN with spectrum width $\Delta
E$ is injected into a hopper like cavity (HC), which is coupled to RC
as shown in Fig. \ref{slope-cavity}. From the uncertainty relation
$\Delta E \Delta t \sim \hbar$ and neutron energy $E = mv^2/2$, we
cant get the coherent length of a single neutron in the monochromatic
beam is $L_c = v \Delta t \sim \hbar/(m\Delta v)$. We propose the
following condition on the monochromaticity of UCN
\begin{eqnarray}
m(\Delta v)^2 < (E_{0,1} - E_{0,0}) \; . \label{require-delta-v}
\end{eqnarray}
Eq.(\ref{require-delta-v}) has two consequences: 1. There would not
be enough kinetic energy for the neutron to knock another neutron to
the excited states; 2. The relative displacement of two neutrons
within the time interval $\tau_g$ is less than $L_c$. Now suppose one
neutron approaches another neutron in the RC. When the distance
between them is within the coherent length of $L_c$, the two neutrons
become indistinguishable. As there is not enough energy to overcome
the gap between $E_{0,1}$ and $E_{0,0}$ from any known types of
interactions (see appendix C), the spatial wave functions of both
neutrons will stay in the ground state and we have
\begin{eqnarray}
\Psi^{(12)}(x_1,y_1; x_2,y_2) = \Psi_{0,0}(x_1,y_1)\Psi_{0,0}(x_2,y_2) \; ,
\end{eqnarray}
where the indexes of the coordinates stand for the two neutrons. Due
to the Pauli exclusive principles of the fermions, the spin wave
function of the two neutrons then must be adjusted to the singlet
state
\begin{eqnarray}
\chi^{(12)} = \frac{1}{\sqrt{2}}(|+-\rangle -|-+\rangle) \; ,
\end{eqnarray}
which is a spin entangled state.

In the following, we shall give an estimation for the practical
experimental realization of our scheme. We take the UCNs with $v = 5$
ms$^{-1}$ for numerical evaluation hereafter \cite{N-velocity}. From
the the condition of Eq.(\ref{require-delta-v}), the monochromatic
UCN beam is characterized by
\begin{eqnarray}
\Delta \lambda/\lambda = \Delta v/ v < 10^{-3} \; , \label{monochramticity}
\end{eqnarray}
This monochromaticity could be achieved via a Drabkin energy filter
\cite{spatial-spin-resonance} which is composed of a polarizer,
spatial spin resonance (SSR) units, and a polarization analyzer.
Using the superconducting solenoid-polarizer \cite{SC-polarizer}, the
UCNs with different polarizations (parallel or antiparallel to
magnetic field) have contrary behaviors: one polarization passes
unhindered through the solenoid while the other polarization reflects
backward. The degree of polarization of UCN can reach the level of
$100\%$ \cite{Ultra-Neutron-SC-Polarizer}. The spatial spin resonance
units only flip the spin of neutrons with specific wavelength
(velocity) \cite{spatial-spin-resonance}. The polarization analyzer
selects the spin flipped neutrons which now have the monochromatic
wavelength (velocity). A precision of $\Delta \lambda /\lambda \simeq
10^{-2}$ has already been reached in \cite{SSR-JPARC} for neutrons
with $\lambda = 5\ \mathrm{\AA}$. As the resolution of the $\lambda$
is inversely proportional to the number of SSR units that neutrons
pass through, an improvement to less than $10^{-3}$ is technically
straight forward \cite{design-ssr-11}.

After the monochromatization, the UCN beam is then guided into the RC
via HC, see Fig.\ref{slope-cavity}. The area of the entrance of HC is
calculated to be $100\, \mathrm{cm}^{2}$. The collimation of the
monochromatic beam should be $v_{x,y}/v_z < 0.2$. That is $v_{x,y}<1
\text{ms}^{-1}$, and the hight that the neutrons climb is less than
$v_{x,y}^2/(2g)\sim 0.05\text{m}$. Thus all the UCNs that enter the
HC can get into the RC. There, the number of UCNs which are within
coherent length $L_c$ is $n_c = 10 \mathrm{cm} \times 10 \mathrm{cm}
\times L_c \times \rho_{\text{UCN}}$ where $\rho_{\text{UCN}}$ is the
number density of UCN at the entrance of HC. In $t$ seconds there
will be
\begin{eqnarray}
N = n_c^2 \times \frac{vt}{L_c} =vtL_{c} \rho^2_{\text{UCN}}\cdot (10^{4}\text{cm}^4)
\label{density-estimation}
\end{eqnarray}
pairs of UCNs that are within the coherent length. Taking the value
of $L_c \sim \hbar/(m \Delta v)\sim 7\times 10^{-4}\text{cm}$ into
Eq.(\ref{density-estimation}), we get $ N \approx 3.5\times
10^{3}\text{cm}^6\text{s}^{-1}\cdot t \rho_{\text{UCN}}^2$. The
polarized UCN density $\rho_{\mathrm{UCN}} = 5\, \mathrm{cm}^{-3}$
has already been obtained \cite{Ultra-Neutron-SC-Polarizer}, the
monochromatization can further reduce this density by an order of
$10^{-3}$ (estimated from Eq.(\ref{monochramticity})). Simple
calculation shows that the production of one spin entangled UCN pair
needs about 12 seconds, that is $3.5\times 10^3\cdot 12 \cdot
(5\times 10^{-3})^2 =1.05$ pairs. If we adopt $\rho_{\mathrm{UCN}} =
10 \, \mathrm{cm}^{-3}$ for polarized UCNs then the same productivity
consumes only three seconds.

Finally, the spin entangled UCN pairs appear at the exit of RC. To
pick the entangled pairs from the stream of neutrons, we can use the
time-of-flight measurements which guarantee that the detected two
neutrons are originally within the coherent length in RC. As the
entangled neutron pair leaves the RC, the two neutrons will move back
to back in the $x-y$ plane due to the conservation of momentum. The
out going entangled neutron pairs can be guided to sufficient long
distance as the depolarization effect of UCN in collision with
materials is quite low, $\sim 10^{-5}$ per collision
\cite{depolarization-UCN}. The spin polarization analyzers can be
applied to the well separated entangled neutrons, verifying their
nonlocal correlations.

In conclusion, it is demonstrated that through a particular type of
RC two neutrons immersed in the earth's gravitational field will
entangle with each other. The predicted production can be 1 entangled
UCN pair in very few seconds. This enables us to test the quantum
nonlocality involving the gravity, the only fundamental interaction
of nature which has not yet been touched in practical entanglement
generation so far. Due to the high detection efficiency of massive
particle and the manageable large spatial separation between two
neutrons (mean life of neutron at rest is $\tau \sim 885.7$s
\cite{PDG}), the proposed scheme also provides a simple and practical
way for the implementation of nonlocality test of quantum
entanglement and statistics in gravitational field, while a more
conclusive test of local hidden variables theory would also be
expected. Most importantly, our experimental scheme has been proved
to be very feasible with the current technique, and a practical
realization can be predicted in the very near future.

\noindent {\bf Acknowledgments}

This work was supported in part by the National Natural Science
Foundation of China(NSFC) under the grants 10935012, 10821063 and
11175249.

\newpage

\noindent {\bf \large Appendix}

\noindent {\bf A. Separation of the variables}

Considering the potential of equation (\ref{V-Gravity-Potential}) in
the region of $x\geq 0$, and $y\geq 0$, equation
(\ref{N-wave-equation}) can be expressed as
\begin{eqnarray}
\left(-\frac{\hbar^2}{2m} \frac{\partial^2}{\partial x^2} +
\frac{mgx}{\sqrt{2}} \right) \Psi(x,y) +  \left( -\frac{\hbar^2}{2m}
\frac{\partial^2}{\partial y^2} +
\frac{mgy}{\sqrt{2}}\right)\Psi(x,y) = E \Psi(x,y)
\end{eqnarray}
Defining $\Psi(x,y) = \psi(x)\psi(y)$, the separation of variables,
and $E = E_x + E_y$, the above equation can then be expressed as
\begin{eqnarray}
\left\{\begin{array}{l} \left(\displaystyle -\frac{\hbar^2}{2m}
\frac{\partial^2}{\partial x^2} + \frac{mgx}{\sqrt{2}} \right)
\psi(x) = E_x \psi(x) \\ \\ \left(\displaystyle -\frac{\hbar^2}{2m}
\frac{\partial^2}{\partial y^2} + \frac{mgy}{\sqrt{2}}\right)\psi(y)
= E_y \psi(y)
\end{array}\right. \; .
\end{eqnarray}
These two equations can be solved in the similar way as equation
(\ref{1-D-wave-function}). The solutions are
\begin{eqnarray}
\psi_{n}(x) = \mathcal{N}_n\,\mathrm{Ai}\left(x/l_0 -
E_{n}/\varepsilon_0 \right)\; ,\; \psi_{m}(y) =
\mathcal{N}_m\,\mathrm{Ai}\left(y/l_0 - E_{m}/\varepsilon_0 \right)
\; , \nonumber
\end{eqnarray}
where  $\mathrm{Ai}$ are Airy functions; $n,m \in \{0, 1, 2, \cdots
\}$; $\mathcal{N}_n$ is the normalization constant; $l_0$,
$\varepsilon_0$ are the characteristic length and energy defined as
$l_0 = \hbar^{2/3}/(\sqrt{2}m^2g)^{1/3}$, $\varepsilon_0=
\sqrt[3]{mg^2\hbar^2/4}$.

\begin{figure}[ht]
\centering \scalebox{0.65}{\includegraphics{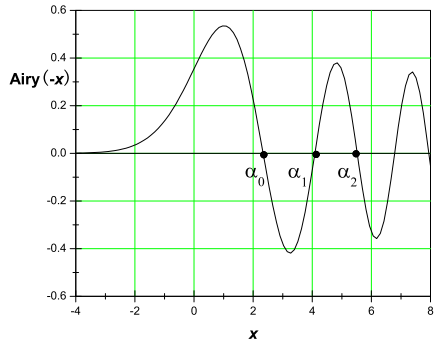}}
\caption{\small The zeros of Airy function: $\mathrm{Ai}(-\alpha_n)
= 0$, $n\in \{0, 1, 2, \cdots\}$. } \label{method-fig-airy}
\end{figure}

The eigenvalues of energy can be obtained by imposing the boundary
condition $\psi_n(0) = \mathcal{N}_n \mathrm{Ai}(-E_n/\varepsilon_0)
= 0$. We can get $E_{n} = \alpha_n \varepsilon_0$, where $\alpha_n$
is the $n$th zero of Airy function, see Figure
\ref{method-fig-airy}.

\noindent {\bf B. Bouncing in the cavity}

In the transection plane of the cavity, the ground state in the
gravitational potential is $\Psi_{0,0}(x,y) = \psi_{0}(x)
\psi_{0}(y)$. Inputting the numerical values $E_0$, $\mathcal{N}_0$
into the wave function we have
\begin{eqnarray}
\Psi_{0,0}(x,y) & \simeq & \frac{25}{81}\mathrm{Ai}(x/6.59-2.34)
\cdot \mathrm{Ai}(y/6.59-2.34) \; , \\ E_{0,0} & = & 2.22\cdot
10^{-12}\ \mathrm{eV} \; ,
\end{eqnarray}
where $x$, $y$ are in unit of $\mu \mathrm{m} = 10^{-6}\mathrm{m}$.
In the ground state, $\langle x \rangle = \langle y\rangle \simeq
10.29\ \mu\mathrm{m}$, $\langle x^2 \rangle= \langle y^2 \rangle
\simeq 126.89\ (\mu\mathrm{m})^2 $, then $ \Delta x = \Delta y =
4.59\ \mu\mathrm{m}$. Due to the Heisenberg uncertainty relation
$\Delta x \Delta p_x \geq \hbar/2$ and $p_x = mv_x$, we have $\Delta
v_x \geq 6.86\cdot 10^{-3}\ \mathrm{ms}^{-1}$. The maximum velocity
in $x$ axis is $v_{x\mathrm{max}} = \sqrt{2E_0/m} \simeq 1.46 \cdot
10^{-2}\ \mathrm{ms}^{-1}$. The average velocity $\overline{v}_x$ in
$x$ axis satisfies
\begin{eqnarray}
\overline{v}_x = \frac{v_{x\mathrm{max}}}{2}\lesssim \Delta v_x\; .
\label{variance-vx}
\end{eqnarray}
The number of bouncing times of the neutron in the cavity is $n =
\displaystyle \frac{T \cdot \overline{v}_x}{2 l_x}$, where $l_x$ is
the side length of the transection plane. The uncertainty of the
bouncing times $n$ arises in regard of the variance $\Delta v_x$,
and can be expressed as $\Delta n = \displaystyle \frac{T \Delta
v_x}{2 l_x} \gtrsim n$, which tells $\Delta n \geq 1$. Consequently,
it is not distinguishable whether the neutron bounces in the cavity
with odd or even number of times.

\noindent {\bf C. Magnetic dipole-dipole interaction}


The energy of magnetic dipole-dipole interaction between two neutrons
is
\begin{eqnarray}
U = \frac{\mu_0}{4\pi}\frac{1}{|\vec{r}_{12}|^3}[ \vec{\mu}_1\cdot\vec{\mu}_2
-3(\vec{\mu}_1\cdot \vec{n})(\vec{\mu}_2\cdot \vec{n})] \; ,
\end{eqnarray}
where $\mu_0$ is permeability of free space, $\vec{\mu}_{1,2}$ are
the magnetic moment of the two neutrons, $|\vec{r}_{12}|$ is the
distance between two neutrons and $\vec{n} =
\vec{r}_{12}/|\vec{r}_{12}|$. The nearest distance between the two
neutrons happens when they are in the same plane of transection of
RC, then the distance can be expressed as
\begin{eqnarray}
r_{12}^2 = |\vec{r}_{12}|^2 = (x_1-x_2)^2 + (y_1-y_2)^2
\end{eqnarray}
the energy of magnetic dipole-dipole interaction between these two
spin parallel neutrons (along $z$ direction) now becomes
\begin{eqnarray}
U = \frac{\mu_0}{4\pi}\frac{\vec{\mu}_1\cdot\vec{\mu}_2}{|\vec{r}_{12}|^3} \; .
\end{eqnarray}
The expectation value $\langle r_{12}^2 \rangle$ can be evaluated
with the spatial wave function $\Psi_{12}(x_1,y_1; x_2,y_2)$ which is
$\overline{r_{12}} = \sqrt{\langle r_{12}^2 \rangle} \sim 9.2
\mu\mathrm{m}$. At this distance the energy of spin-spin coupling of
the two neutrons is roughly $ U \sim 10^{-23}\mathrm{eV}$. This is
negligibly small compare to the energy gap $E_{0,1}-E_{0,0}\sim
10^{-12}\mathrm{eV}$.

\end{document}